\documentclass[pre,superscriptaddress]{revtex4}
\usepackage{amsmath}
\usepackage{epsfig}

\begin{document}

\title{Homogeneous nucleation of colloidal melts under the influence
  of shearing fields}

\author{Ronald Blaak}
\affiliation{Institut f\"ur Theoretische Physik II,
  Heinrich-Heine-Universit\"at, Universit\"atsstra{\ss}e 1, D-40225
  D\"usseldorf, Germany} 

\author{Stefan Auer}
\affiliation{Department of Chemistry, Cambridge University, Lensfield
  Road, Cambridge, CB2 1EW, United Kingdom} 

\author{Daan Frenkel}
\affiliation{FOM Institute for Atomic and Molecular Physics, Kruislaan
  407, 1098 SJ Amsterdam, The Netherlands} 

\author{Hartmut L\"owen}
\affiliation{Institut f\"ur Theoretische Physik II,
  Heinrich-Heine-Universit\"at, Universit\"atsstra{\ss}e 1, D-40225
  D\"usseldorf, Germany} 


\begin{abstract}
We study the effect of shear flow on homogeneous crystal nucleation,
using Brownian Dynamics simulations in combination with an umbrella
sampling like technique. The symmetry breaking due to shear results in
anisotropic radial distribution functions. The homogeneous shear rate
suppresses crystal nucleation and leads to an increase of the size of
the critical nucleus. These observations can be described by a 
simple, phenomenological extension of classical nucleation theory. In
addition, we find that nuclei have a preferential orientation with
respect to the direction of shear. On average the longest dimension of
a nucleus is along the vorticity direction, while the shortest
dimension is preferably perpendicular to that and slightly tilted with
respect to the gradient direction. 
\end{abstract}


\maketitle 

\section{Introduction}
It is well known that fluids can be cooled to temperatures below
coexistence temperature of the liquid and solid phase. Density
fluctuations in this super-cooled regime lead to the continuous
formation of small crystalline nuclei. Even though the crystalline
phase is the thermodynamical stable phase, a super-cooled system can
remain for longer times liquid-like. The reason lies in the fact that
there is a counter acting force to crystallisation which finds its
origin in the interface that needs to be formed between small
crystallites and the surrounding liquid. This results in a nucleation
barrier which needs to be overcome in order to solidify the system. This
intuitive picture forms the basis of what at present is known as
classical nucleation theory~\cite{Volmer:1926ZPC,Becker:1935AnP}. 

Understanding the underlying principles of nucleation and the growth
to macroscopic sizes finds its applications in many areas,
e.g. protein crystallisation and
metallurgy~\cite{Kelton:1991,Galkin:2000PNAS,Shi:1995APL}. The first
experimental studies of crystal nucleation are light scattering
experiments~\cite{Schatzel:1993PRE,Harland:1997PRE,Sinn:2001PCPS} in
hard sphere colloids, which form a well understood model
system~\cite{Pusey:1991}. Crystallisation rates were measured by 
Gasser {\em et al.} using confocal microscopy in weakly
charged colloids~\cite{Gasser:2001SC}.
Recently computer simulations have been used to explore nucleation 
phenomena and by using special simulation techniques it was possible
to obtain quantitative results on the absolute rate of crystal
nucleation in colloidal suspensions~\cite{Auer:2001NAT1,Auer:2002JPCM}
and compare them with experiments.

In the present work we focus on nucleation outside equilibrium,
i.e. in the presence of a homogeneous shear field. Experiments on
crystallisation under shear suggest two different scenarios. In some
cases one observes that crystallisation is enhanced by the presence of
shear~\cite{Ackerson:1988PRL,Yan:1994PA,Haw:1998PRE,Amos:2000PRE}.
This can, at least qualitatively, be understood, by the idea that
shear might facilitate the formation of layers in the system, making
it easier to crystallise. Other experiments report that
crystallisation is
suppressed~\cite{Palberg:1995JCP,Okubo:1999JCIS}. In the later case
one could imagine that the shear will de-attach particles from newly
born crystallites. Both effects are also found in heterogenous
nucleation under shear in single or double wall confinement. For low
and moderate shear rates two different types of layering sliding are
observed, while for high shear rates the layers are found to
destabilise~\cite{Stipp:2004JPCM}. 

A priory it is not clear which of the both scenarios will dominate and
how it depends on the external conditions. In simulations by Butler
and Harrowell~\cite{Butler:1995PRE} for the crystallisation kinetics
of particles with Yukawa interaction subjected to shear rates, it was
observed that crystallisation is suppressed by shear. In the present
work we will focus on nucleation in its bare bones and extract
information on the properties of nuclei that are formed in a
super-cooled melt. A preliminary account of our simulation results was
already published~\cite{Blaak:2004PRL}. We include here further data for
pair correlations, the critical nucleus structure and orientation, and
describe the simulation technique in more detail. 

The remainder of this paper is organised as follows. In
Sec.~\ref{sec:model} we introduce the model and explain the
combination of different techniques we used for our simulations. The
simulation results for the pair-correlation functions and the size
distribution function of the nuclei are shown in
Sec.~\ref{sec:sim}. In Sec.~\ref{sec:cnt} we make a simple,
phenomenological extension of classical nucleation theory and compare
this with the results obtained from the simulations. A more detailed
analysis of the shape and relative orientation with respect to the
shear direction is performed in Sec.~\ref{sec:nucleus}. We finish in
Sec.\ref{sec:conclusion} with a summary and discussion of our main
results. 

\section{The model}
\label{sec:model}
In order to simulate homogeneous crystal nucleation under shear, we
have chosen to consider the simple model of charge-stabilised
colloidal suspensions subjected to a linear shear flow. Due to
screening of the micro-ions, these particles mutually interact
effectively via a repulsive Yukawa potential~\cite{Pusey:1991} 
\begin{equation}
\label{eq:pairpot}
V(r) = \epsilon \frac{e^{-\kappa r}}{\kappa r},
\end{equation}
where $r$ is the distance between the centres of the colloidal
particles. The magnitude of screening is expressed by the inverse
screening length $\kappa$ and the strength of the interaction is
denoted by $\epsilon$. 

In addition to the normal inter-particle interaction, we also want to
apply shear to the system. This means that we are dealing with a
non-equilibrium situation, hence we need to make use of dynamical
simulations rather than simulations of the Monte Carlo type. For low
volume fractions of charged suspensions one can neglect the
hydrodynamic interactions between the colloids, which enables us to
apply Brownian
Dynamics~\cite{Book:Allen-Tildesley,Chakrabarti:1994PRE}. In this
over-damped motion, it is assumed that particles move at constant
velocity proportional to the force due to other particles, and that
the solvent exerts a random force with a Gaussian distribution. The
steady shear rate $\dot{\gamma}$ imposed on the system, is assumed to
generate a linear velocity gradient in the solvent, which in turn
results in a linear force field for the colloids. The combination of
Brownian Dynamics and shear results in the following equations of
motion 
\begin{equation}
\label{eq:step}
\vec{r}_i(t+\delta t) = \vec{r}_i(t) + \delta t
\frac{\vec{f}_i(t)}{\xi} + \delta \vec{r}^G + \delta t \dot{\gamma}
y_i(t) \hat{x} \;. 
\end{equation}
Here $\vec{r}_i(t)=(x_i(t),y_i(t),z_i(t))$ is the position of the
$i$th colloidal particle at time $t$. In a small time interval $\delta
t$ this particle moves under influence of the sum of the conservative
forces $\vec{f}_i(t)$ arising from the pair interaction
(\ref{eq:pairpot}) of particle $i$ with the neighbouring
particles. The friction constant $\xi$ with the solvent is related to
the short-time diffusion constant $D$ by $\xi = k_B T /D$. The
stochastic displacements are independently drawn from a Gaussian
distribution with zero mean and variance $\langle (\delta
r_{i\alpha}^G)^2\rangle = 2 D \delta t$, where $\alpha$ stands for one
of the Cartesian components. The last term in Eq.~(\ref{eq:step})
represents the applied shear in the $x$-direction, and imposes an
explicit linear flow field. 

All simulations reported here are done for a system of 3375 colloidal
particles and the interaction strength is fixed to $\epsilon = 1.48
\times 10^4 k_B T$. For practical purposes the interaction potential
is truncated at the cut-off distance $r_c = 10/\kappa$, by shifting
the potential to zero. The inter-particle forces are not effected by
this procedure. For simplicity we use a cubic simulation box to which
we apply the shifted periodic boundary conditions. These were
introduced by Lees and Edwards~\cite{Lees:1972JPCS} and are required
to deal with the presence of shear. 

The simple combination of Brownian Dynamics with a linear imposed
velocity field and Lees-Edwards boundary conditions, forms the heart
of the simulations. This ensures that the starting configurations
follow the proper equations of motions. To facilitate the connection
with Gibbs free energies, however, it is desirable to do the
simulations at constant osmotic pressure. Hereto we have added volume
moves as normally used in equilibrium Monte Carlo simulations of the
isobaric ensemble. In practice this means that after a large number of
Brownian Dynamics steps, the simulation box expands or shrinks
isotropically. The difference in potential energy due to this volume
move, is used in order to accept or reject this move. The value of the
applied pressure is chosen such that the bulk behaviour of the system
is in the super-cooled regime. Therefore the prepared system will
remain liquid, even though the crystalline phase is more stable. Due
to fluctuations inside such a liquid, small nuclei are continuously
formed and dissolved. Only once a sufficiently large nucleus is
formed, the critical nucleus size, the growth of the nucleus will
dominate and can lead to a full crystallisation of the system. 

The quantity of interest of our simulations is the cluster size
distribution function $P(n)$, which describes the probability of
finding a nucleus of $n$ particles. This is a well defined object,
even in the non-equilibrium situation of the presence of shear,
provided that after a sufficient long simulation time the system will
reach a steady state situation. To this end we need to analyse the
configurations obtained by the simulations and detect all the nuclei
in the liquid. This can be done with the aid of bond-orientational
order parameters introduced by Steinhardt {\em
et~al.}~\cite{Steinhardt:1983PRB} and applied to study crystal
nucleation by Frenkel and
coworkers~\cite{Duijneveldt:1992JCP,Wolde:1995PRL,Auer:2004JCP}.
 All particles within a short distance of the particle of
interest are determined, and the vectors connecting them are resolved
in polar angles. These can be used to determine rotationally invariant
order parameters, which characterise the neighbourhood of a particle, and
allow us to distinguish a local fluid environment from a crystalline
environment. It is even possible to distinguish between different
crystalline structures. Solid particles that are in each others
neighbourhood belong to the same cluster. In this manner we obtain the
total number of clusters in the system and each of their sizes.

By performing long simulations, in which between $10^6$ and $10^7$
Brownian Dynamics steps are made, we can now
measure the probability size distribution function $P(n)$ in a simple
way. After 20 Brownian Dynamics steps and possibly a volume move
(on average one per hundred BD steps), we analyse the
configuration in determining the size and the number of clusters
present in the liquid. These are used to make a histogram of the
possible cluster sizes, and the frequency is proportional to the
probability of observing such a cluster size. Note that also in the
presence of shear, and therefore in non-equilibrium, this procedure
can be followed.

Nucleation, however, is a rare process and unless the system has a
large nucleation rate~\cite{OMalley:2003PRL}, no sufficient statistics
can be obtained by merely the technique described above. The solution
to this problem is the usage of umbrella
sampling~\cite{Torrie:1974CPL}. This technique stems from equilibrium
Monte Carlo simulations and introduces an additional potential to the
system. Such a bias should depend on a relevant reaction coordinate of
the system under consideration and is able to restrain the system in
meta-stable situation. The traditional interpretation is that the bias 
is used as an external potential, leading to a biased ensemble. It
therefore might seem somewhat surprising that the technique can be
used even in non-equilibrium situations. The reason for this apparent
ambiguity, is that umbrella sampling can be considered as a plain
mathematical trick.

In order to measure the steady state probability distribution $P(n)$,
we need to sample the relative probabilities of $P(n+1)/P(n)$, which
is achieved by growing paths by Brownian dynamics simulation from one
cluster size to another. In principle this could be done by starting
with a chosen cluster-size $n$ and count the number of times the
cluster grows or shrinks with one particle or remains at the same size
in a small interval of time, and average afterwards over many
different realizations of the initial configuration. The number of
counts of a given size $n$ is then proportional to $P(n)$. The obvious
problem is that below the critical nucleus size, the cluster will
mainly shrink, and beyond the critical size it will mainly grow. This
would be alright, but causes a different problem, i.e. how to obtain
many configurations of a given size which have a very low
probability. 

This problem is circumvented by using the bias. For each simulation we
use a fixed biasing function based on some preferred cluster-size
$n_0$, which has the simple form 
\begin{equation}
\label{eq:bias}
V_{bias}(n) = \frac{1}{2} \alpha (n - n_0)^2,
\end{equation}
with $\alpha$ is a measure for the strength of restrainment. We start
with a given configuration and follow the normal Brownian Dynamics as
described above. After a fixed time interval we measure the size $n$
of the largest cluster. We now continue with either the final
configuration or the previous starting configuration as the new
initial configuration. Only this is determined by the bias we
introduce, it therefore has no effect on the dynamics of the system
and is just an aid to generate new initial configurations. By choosing
the bias appropriate to a harmonic-like form, we ensure that the
probability of growing a bigger or shrinking to a smaller cluster with
respect to $n_0$ is approximately the same. The count of the number of
clusters of a given size, is now proportional to the ratio of $P(n)$
and $P_{bias}(n)$ of the known bias we impose. By performing
simulations for different preferred sizes $n_0$, we obtain successive
parts of the function $P(n)$ which can be combined to form the
complete function. 

\section{Simulation results}
\label{sec:sim}
The pressures at which simulations are performed are such that the
equilibrium bulk phase without shear is a face-centered-cubic
crystalline phase~\cite{Auer:2002JPCM,Hamaguchi:1997PRE}. Hence the
liquid state in which we prepare the system is only
meta-stable. Nevertheless, the pressure is still small enough such
that spontaneous crystallisation is not observed within the duration
of the simulation and even in the absence of the bias the system
remains liquid and only forms nuclei smaller than the critical nucleus
size. 

\begin{figure}[h]
\epsfig{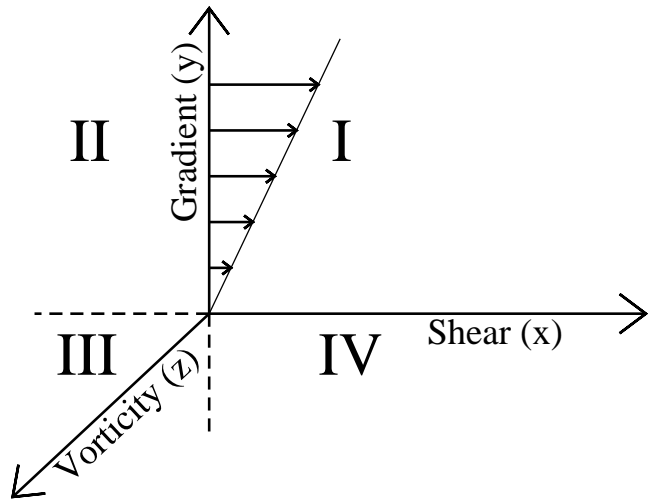}
\caption[a]{Schematic drawing of the different directions in the
  problem. The roman numbers label the different quadrants of space.} 
\label{Fig:Schematic}
\end{figure}

One of the consequences of the presence of shear is that this
introduces a symmetry breaking in the fluid. This can be seen easily
in the 2-particle distribution function
$g^{(2)}(\vec{r_1},\vec{r_2})$. Whereas in bulk-equilibrium, this
function is only dependent on the absolute distance
$|\vec{r_1}-\vec{r_2}|$, in the presence of shear the function will
also depend on the relative direction. In general the function can be
expanded in a series of modified spherical harmonics
$C_{l,m}(\theta,\phi)$ with distant dependent coefficients
$g_{l,m}(r)$ 
\begin{equation}
g(r,\theta,\phi) = \sum_{l=0}^{\infty} \sum_{m=-l}^{l} \frac{2 l +
  1}{4 \pi} g_{l,m}(r) C_{l,m}(\theta,\phi), 
\end{equation}
where $\theta$ is the angle with respect to the vorticity direction,
i.e. $\cos \theta = \hat{r} \cdot \hat{z}$, and $\phi$ the angle with
the shear direction, i.e. $\cos \phi = \hat{r} \cdot \hat{x}/\sqrt{1 -
 (\hat{r} \cdot \hat{z})^2}$ (See Fig.~\ref{Fig:Schematic}). Note
that without the symmetry breaking due to shear, only the isotropic
contribution remains, i.e. the usual radial distribution function
$g(r)\equiv g_{0,0}(r)$. 

The distance dependent coefficients follow directly from the
orthogonality relation of the spherical harmonics 
\begin{equation}
g_{l,m}(r) = \int d \cos \theta \int d \phi ~ g(r,\theta,\phi)
C_{l,m}^*(\theta,\phi) 
\end{equation}
and can easily be obtained from simulations. Due to symmetries present
in the system and the suitable chosen reference frame, only functions
with both indices $l$ and $m$ even have non-zero contributions. It
turns out that the first order correction to the radial distribution
function is dominated by the imaginary part $\Im(g_{2,2}(r))$ of the
function $g_{2,2}(r)$. 

\begin{figure}[ht]
\epsfig{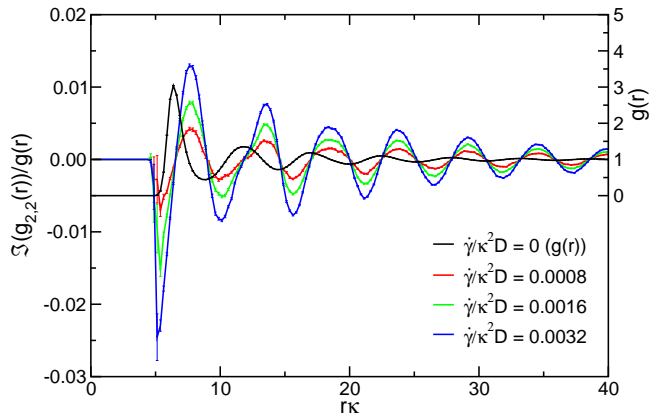}
\caption[a]{The imaginary part of the function $g_{2,2}(r)$ as
  function of the distance for fixed pressure $\beta P/\kappa^3 =
  0.240$ and different shear rates. For comparison the radial
  distribution function $g(r)$ is included.} 
\label{Fig:radialC22}
\end{figure}

In Fig.~\ref{Fig:radialC22} we show the ratio of $\Im(g_{2,2}(r))$ and
the function $g_{0,0}(r)$ as obtained from simulations without bias
and for different shear-rates. The corresponding function, which was
statistically averaged, is $\Im(C_{2,2}) = -\frac{1}{4} \sqrt{3}
\sin^2 \theta \sin 2 \phi$. This quantity is able to distinguish
whether a particle is preferably found in the first (third) or in the
second (fourth) quadrant of space provided another particle is in the
origin (See Fig.~\ref{Fig:Schematic}). The signal is phase-shifted
with respect to the oscillations in the radial distribution function,
such that particle distances in a peak of a distribution function and
located in the first quadrant are typically closer than those
particles found in the second quadrant. It can also be observed that
the amplitude grows linearly with the shear rate. The distribution
functions $g_{0,0}(r)$ are, within the accuracy of the simulations,
identical for these moderate shear rates. 

In order to obtain the probability function $P(n)$ of observing a
cluster of $n$ particles, we perform simulations for fixed pressure
and shear-rate. The values of the preferred size $n_0$ in the bias
(\ref{eq:bias}) is chosen in multiples of 10 and the strength $\alpha
= 0.15$. This ensures that the window of observed cluster sizes are
sufficiently wide to produce overlap between neighbouring
windows. When the bias is included, we do not only find a single big
nucleus, but also small nuclei can be observed. It suffices, however,
only to count the largest cluster, as it are those fluctuations in
size that determine the probability function $P(n)$. The overlap
between neighbouring windows is used to match successive parts of
$P(n)$. The only exception is the first part of the function, where
small nuclei are important. For that region an unbiased simulation is
used in which all clusters in the system are used. For a more detailed
discussion on how the curve is obtained we refer the reader
to~\cite{Auer:2004JCP}. 

\begin{figure}[h]
\epsfig{figure=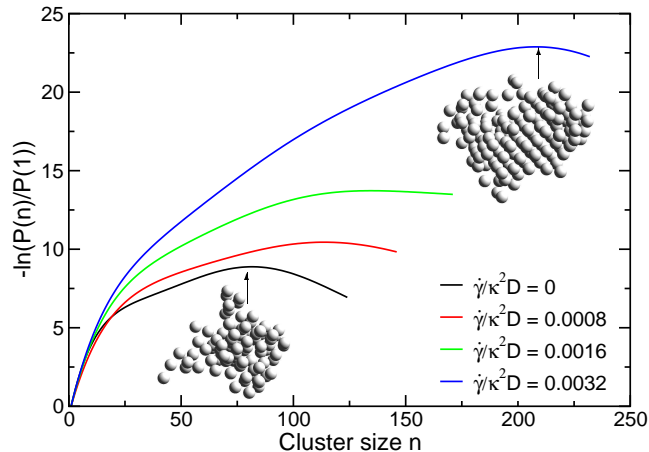,width=8.5cm,angle=0}
\caption[a]{Negative logarithm of the probability $P(n)$ of finding a
  cluster of $n$ solid-like particles normalised by $P(1)$ for
  pressure $\beta P/\kappa^3 = 0.24$ and different applied shear
  rates, from bottom to top $\dot{\gamma}/(\kappa^2 D)=0$, $0.8 \times
  10^{-3}$, $1.6 \times 10^{-3}$, and $3.2 \times 10^{-3}$. The insets
  show typical snapshots of critical nuclei for the largest shear rate
  and the zero shear case.} 
\label{Fig:LogP}
\end{figure}

The results of the simulations for a single pressure and several
shear-rates are shown in Fig.~\ref{Fig:LogP}. It shows the negative
logarithm of the relative probability function $P(n)$ normalised with
the probability $P(1)$ of finding a cluster of only a single
particle. For increasing values of the shear rate, the maximum in the
curves shifts towards larger values and larger cluster sizes. 

\section{Extension of classical nucleation theory}
\label{sec:cnt}
Classical nucleation theory links the probability of finding a cluster
of given size to the Gibbs free energy and allows to interpret a
probability function as shown in Fig.~\ref{Fig:LogP} for the
zero-shear case in terms of nucleation barrier~\cite{Auer:2004JCP}. It
splits the cost in Gibbs free energy $\Delta G$ to create a spherical
nucleus in a liquid in two contributions 
\begin{equation}
\label{eq:cnt}
\Delta G = - \frac{4}{3} \pi R^3 \rho_S | \Delta \mu | + 4 \pi R^2 \gamma_{SL}.
\end{equation}
On the one hand there is a gain in energy proportional to the volume
of the nucleus with radius $R$ due to the difference in chemical
potential $\Delta \mu$ between the solid with density $\rho_S$ and the
liquid phase. On the other hand we have a loss in energy, since an
interface between the solid nucleus and surrounding liquid needs to be
formed, described by $\gamma_{SL}$ the interfacial free energy. 

It can be shown that the free energy cost of creating a nucleus is
related to the probability function $P(n)$, and if the appropriate
correction is included~\cite{Auer:2004JCP}, the height of the maximum
in the curve corresponds to the nucleation barrier, while the
corresponding cluster size is the size of the critical nucleus. The
actual shape of the barrier, as found in simulations, will depend on
the way in which a cluster is defined. This will therefore also
influence the number of particles in the critical nucleus. The
nucleation barrier itself, however is much less dependent on the
cluster definition. In the presence of shear the curves have a similar
shape. Also here we can identify the top with a critical nucleus size,
since for smaller clusters there is tendency to shrink, while for
larger clusters the growth will dominate. The height of the top is a
measure for the likeliness that such a cluster will be formed and is
therefore related to the time one needs to wait in order to observe
such a cluster size. 

It is therefore tempting to interpret also the other probability
functions of Fig.~\ref{Fig:LogP} in terms of nucleation
barriers. However, due to the presence of shear in the system, we are
dealing with a non-equilibrium situation, and although the concept of
a nucleation barrier in this case seems intuitively clear, strictly
speaking it is wrong since the Gibbs free energy is ill defined
outside equilibrium. Surprisingly, the application of statistical
mechanics to non-equilibrium situation can be useful (see
e.g,~\cite{Daivis:2003JCP}) and it is a canonical challenge to check
the validity of carrying over equilibrium concepts into
non-equilibrium physics. With all the principle caveats which one
should have in mind about this, we here will assume that we can extend
the ideas of classical nucleation theory to the present situation by
including shear. 

As is the case in classical nucleation theory, we assume that also in
the presence of shear a spherical nucleus is formed. The two terms in
Eq.~(\ref{eq:cnt}) will therefore remain. It is however reasonable to
expect that the difference in chemical potential $\Delta \mu$ and
interfacial free energy $\gamma_{SL}$ will be affected by the shear,
but that for moderate shear-rates an expansion in powers of the shear
rate for both these quantities about their equilibrium values can be
performed 
\begin{equation}
\label{eq:mugamma}
\begin{split}
\Delta \mu & = \Delta \mu^{(eq)} \left(1 + c_0 \dot{\gamma}^2 + {\cal
 O} (\dot{\gamma}^4) \right) \\ 
\gamma_{SL} & = \gamma_{SL}^{(eq)} \left(1 + \kappa_0 \dot{\gamma}^2 +
 {\cal O} (\dot{\gamma}^4) \right), 
\end{split}
\end{equation}
where due to the invariance of the shear direction only even powers in
the shear rate $\dot{\gamma}$ need to be considered. If we insert the
expansions (\ref{eq:mugamma}) in the expression from classical
nucleation theory (\ref{eq:cnt}) we obtain simple expressions for
$\Delta G^*$, the height of the nucleation barrier and $N^*$, the size
of the critical nucleus 
\begin{equation}
\label{eq:dGN}
\begin{split}
\Delta G^* = \frac{16 \pi \left[ \gamma_{SL}^{(eq)}\right]^3}{3
 \rho_S^2 \left| \Delta \mu^{(eq)} \right|^2} \left[ 1 + (3 \kappa_0
 - 2 c_0) \dot{\gamma}^2 + {\cal O} (\dot{\gamma}^4) \right] \\ 
N^* = \frac{32 \pi \left[ \gamma_{SL}^{(eq)}\right]^3}{3 \rho_S^2
 \left| \Delta \mu^{(eq)} \right|^3} \left[ 1 + (3 \kappa_0 - 3 c_0)
 \dot{\gamma}^2 + {\cal O} (\dot{\gamma}^4) \right], 
\end{split}
\end{equation}
where the corrections with respect to the unsheared case scale to
leading order with $\dot{\gamma}^2$. 

\begin{figure}[ht]
\epsfig{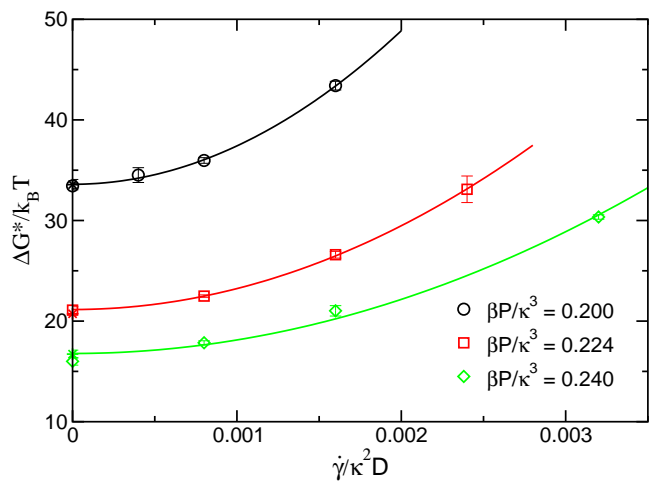}
\caption[a]{The height of the nucleation barrier $\beta \Delta G^*$ as
  a function of the dimensionless shear rate $\dot{\gamma}/\kappa^2 D$
  for different pressures $P$. The solid lines are parabolic fits
  through the data. The stars are data from equilibrium Monte Carlo
  simulations without shear.} 
\label{Fig:dG}
\end{figure}

\begin{figure}[ht]
\epsfig{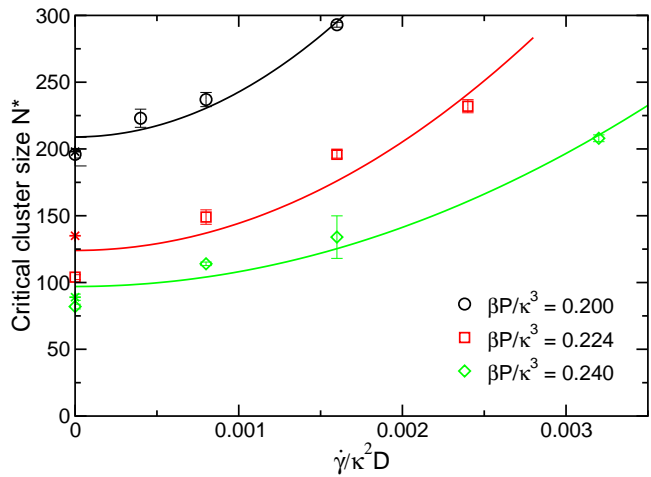}
\caption[a]{The number of particles $N^*$ of the critical nucleus as a
  function of the dimensionless shear rate $\dot{\gamma}/\kappa^2 D$
  and for different pressures $P$. The solid lines are parabolic fits
  through the data. The stars are data from equilibrium Monte Carlo
  simulations without shear.}
\label{Fig:N}
\end{figure}

In Fig.~\ref{Fig:dG} and~\ref{Fig:N} we show the results from our
simulations where we extracted the height of the nucleation barrier
and the size of the critical nucleus for various pressures and shear
rates. For the zero-shear case, the data from equilibrium Monte Carlo
simulations have been included. Parabolic fits through the data
support the quadratic dependency of these quantities on the shear-rate
and also enable us to obtain estimates for the second order
coefficients in Eqn.~\ref{eq:mugamma}, the results of which are
summarised in Table~\ref{Tab:data}. The presence of shear destabilises
the solid as is indicated by the negative value of $c_0$. 

\begin{table}[ht]
\begin{tabular}{c|c|c|c|c}
$\beta P/\kappa^3$ & $\beta \Delta G^{(eq)}$ & $N^{(eq)}$ & $c_0 D^2
  \kappa^4$ & $\kappa_0 D^2\kappa^4$ \\ 
\hline
0.200 & 34 & 209 & -4.8 $\times$ 10$^4$ & 6. $\times$ 10$^3$ \\
0.224 & 21 & 133 & -4.1 $\times$ 10$^4$ & 5. $\times$ 10$^3$ \\
0.240 & 17 &  97 & -3.4 $\times$ 10$^4$ & 4. $\times$ 10$^3$
\end{tabular}
\caption{Numerical data for different pressures $\beta P/\kappa^3$ on
  the equilibrium barrier height $\Delta G^{(eq)}$, critical nucleus
  size $N^{(eq)}$, and second order corrections to the free energy
  difference and interfacial free energy as obtained from the fitted
  simulation data.} 
\label{Tab:data}
\end{table}

The shift of the maximum of the curves in Fig.~\ref{Fig:LogP} as
function of the applied shear rate, is interpreted as an increase in
the nucleation barrier and the critical nucleation size. This
indicates that the interaction of the system under consideration
corresponds to a situation in which crystal nucleation is suppressed
by the presence of shear~\cite{Butler:1995PRE}. In addition, the
application of shear allows us to supercool the system beyond the
point at which in an equilibrium simulation spontaneous
crystallisation appears. Due to system size restrictions, much larger
shear rates could not be accessed, because the increase of the
critical nucleus size would lead to simulation artefacts, i.e. a
percolating cluster. It is imaginable that much larger shear rates
could enhance crystallisation by pre-layering the
system~\cite{Stevens:1993PRE}. 

It is important to realise that the shear rate should not be
considered as a thermodynamic variable~\cite{Daivis:2003JCP}. In fact,
in a recent study of the effect of shear on the location of the
solid-liquid coexistence in a Lennard-Jones system, Butler and
Harrowell found that no purely thermodynamic description of the effect
of shear was possible~\cite{Butler:2003JCP}. Shear directly affects
the transport of particles from the solid to the liquid phase, and
this effect is not thermodynamic. The expansion in
Eq.~\ref{eq:mugamma} is simply a way to represent the effect of shear
as if it were purely thermodynamic. A priory there is no reason to
expect the validity of these assumptions and it is therefore very
surprising that this simple extension of classical nucleation theory
is able to describe our observations. 

\section{The structure and relative orientation of the nucleus}
\label{sec:nucleus}
From an analysis of the bond orientational order parameters it follows
that the structure of the nucleus is predominantly
body-centered-cubic, even though the stable equilibrium phase is the
face-centered-cubic crystal. An intriguing question is whether the
shape and/or orientation of the solid nuclei that are formed in a
super-cooled liquid are influenced by the presence of moderate
shear-rates. Also without shear, a nucleus is in general neither
spherical nor convex. To this end we calculate for each cluster the
moment of inertia tensor. After diagonalisation the principal moments
of inertia are obtained. For a truly spherical object these should be
identical, but due to natural fluctuations this is usually not the
case. 

\begin{figure}[ht]
\epsfig{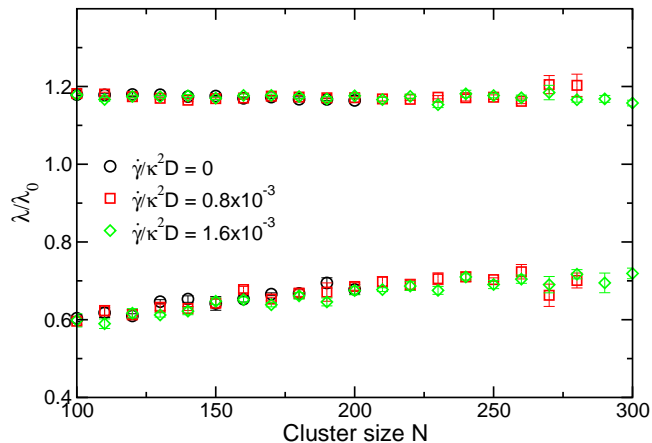}
\caption[a]{The ratio of both the maximum and minimum moments of
  inertia with the medium as function of the number of particles in
  the nucleus and for different shear rates and constant pressure
  $\beta P/\kappa^3 = 0.2$.} 
\label{Fig:mom}
\end{figure}

In Fig.~\ref{Fig:mom} we show the average ratio $\lambda_+/\lambda_0$
and $\lambda_-/\lambda_0$ of the largest ($\lambda_+$) and smallest
($\lambda_-$) moment of inertia with the intermediate ($\lambda_0$)
one as function of the cluster size $n$. The dependence on the cluster
size indicate that larger clusters are slightly more spherical. It is
interesting to observe that the shear rates have no visible influence
on these ratios, which suggest that the shape does not depend strongly
on the applied shear. This is different from the radial distribution
functions we measured in the liquid under shear, as they become
increasingly asymmetric for higher shear rates. 

\begin{figure}[ht]
\epsfig{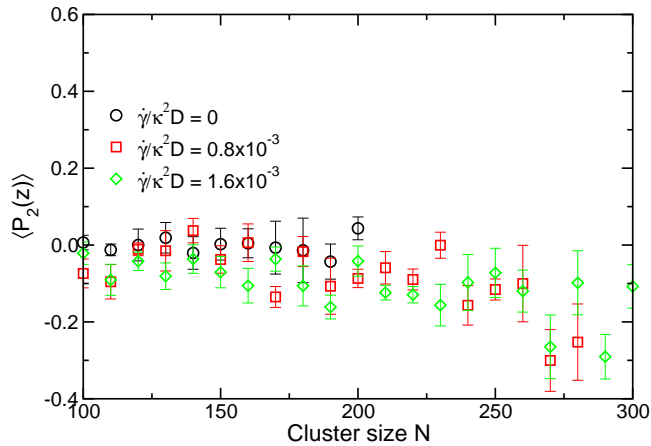}
\caption[a]{The correlation $\langle P_2(z) \rangle$ of the coupling
  between the direction of the major moment of inertia with the
  vorticity direction $\hat{z}$ as function of the size of the nucleus
  and for constant pressure $\beta P/\kappa^3 = 0.2$.} 
\label{Fig:p2z}
\end{figure}

The diagonalisation procedure of the moment of inertia tensor also
enable us to characterise the orientation of the main axes of the
nucleus. Whereas the shape is hardly influenced by the shear, there is
a coupling between the orientation of the nucleus and the direction of
the shear. To quantify this, we calculate the direction of the largest
principal moment (smallest dimension of the nucleus). The
$z$-component of the corresponding unit-vector gives the cosine of the
angle with the vorticity direction. This is weighted with the second
Legendre polynomial $P_2$ in order to take the symmetries into
account. The result is shown in Fig.~\ref{Fig:p2z}, where $\langle
P_2(z) \rangle$ is plotted versus the cluster size. For an isotropic
cluster shape distribution, this correlation function should be zero
as is found for the zero-shear case. In the presence of shear,
however, a negative correlation is found that grows for larger
nuclei. This suggests that the smallest dimension of a nucleus have a
preference for being oriented perpendicular to the vorticity
direction. 

\begin{figure}[ht]
\epsfig{figure=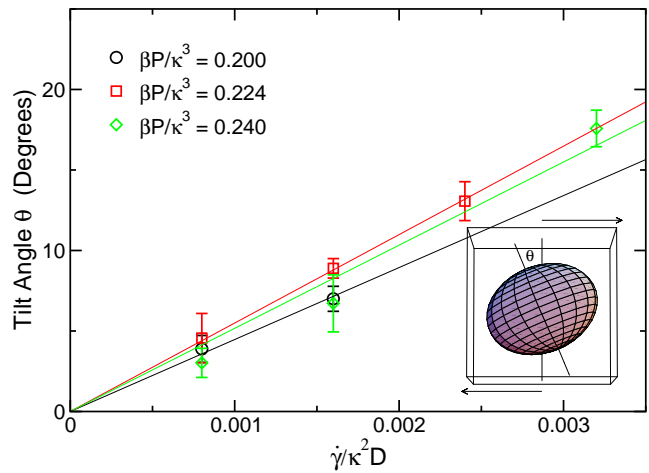,width=8.5cm,angle=0}
\caption[a]{The tilt angle $\theta$ of the principal moment of inertia
  with respect to the $y$-axis. The inset shows a schematic
  representation of the preferred orientation of the nucleus with
  respect to the shear direction indicated by the arrows.} 
\label{Fig:orient}
\end{figure}

In the same manner we can quantify the correlation between the other
directions of the principal moments with axes of the reference
frame. From this analysis it follows that the largest dimension of the
nucleus (smallest moment of inertia) is preferentially along the
vorticity direction, while the smallest dimension prefers to be in the
plane of the gradient and shear direction. Even within this plane, the
distribution of orientations is not isotropic. The measured
correlations, enable us to calculate the average angle made between
the smallest dimension and the gradient direction of which the results
are shown in Fig.~\ref{Fig:orient}. In order to improve the statistics
we have averaged over all clusters containing more than 100 particles
up to the size of the critical nucleus $N^*$. The inset of
Fig.~\ref{Fig:orient} shows a schematic drawing of the preferred
orientation of the nucleus with respect to the reference frame and
indicates the tilt angle $\theta$. Note that the largest dimension of
the nucleus (smallest principal moment) is preferably along the
vorticity direction, i. e. perpendicular to the plane of drawing. 

As can be seen from Fig.~\ref{Fig:orient}, the tilt angle increases
linearly with the applied shear rate $\dot{\gamma}$ and only depends
weakly on the osmotic pressure. Interestingly, a similar tilt occurs
when vesicles with a flexible shape are exposed to a linear shear
flow~\cite{Abkarian:2002PRL}. It is, however, unclear whether this
is coincidental or not. 

\section{Conclusions}
\label{sec:conclusion}
In summary we have studied the influence of shear on the nucleation of
particles with Yukawa interaction. To this end we used a combination
of Brownian dynamic simulations with a sort of umbrella sampling
technique. The main result is that the probability of crystal
nucleation decreases with shear rate, while the critical nucleus size
increases. The shape of nuclei is hardly influenced by the presence of
moderate shear rates, but it will affect their orientation. 

In order to understand the qualitative behaviour of the observations,
we used an empirical extension of classical nucleation theory in which
we carry over simple equilibrium concepts to a situation of
non-equilibrium. A priory there is no reason to expect the validity of
such a procedure, but surprisingly it is capable of describing our
observations and we hope it will stimulate the fundamental research of
non-equilibrium physics. 

The observations made here, could in principle be tested
experimentally, by studying homogeneous crystal nucleation in sheared
colloidal suspensions. If a Poiseuille flow as realized in a capillary
viscometer~\cite{Palberg:1996JP1F} would be used, rather than
homogeneous Couette flow, we should expect crystal nuclei to appear
preferentially in the middle of the flow channel. 

For practical reasons we only considered fluids that are weakly
sheared, for which one might expect that shear-induced layering is
unimportant. It is imaginable that for larger shear such pre-ordering
phenomena could become relevant, and eventually lead to a situation
for which crystal nucleation is
enhanced~\cite{Stevens:1993PRE}. Another important assumption has been
that we ignored hydrodynamic interactions for computational
reasons. Although this is allowed for dilute system, this is in
general not correct. 
Closely related is that the implicit solvent is assumed to have a
homogeneous linear velocity profile and is unaffected by the presence
of the more dense crystalline nucleus. For lower pressures and higher
shear rates this is not correct~\cite{Reguera:2003JCP2}, as the local
transport properties of the liquid can be influenced by the nucleus
and consequently affect the nucleation process.

The method can easily be extended to include oscillatory
shear~\cite{Xue:1989PRA} or to consider heterogeneous
nucleation~\cite{Auer:2003PRL} near a system wall in a  
sheared suspension. The combination of umbrella sampling with a
dynamical simulation method in order to obtain information on rare
events is generally applicable in equilibrium and non-equilibrium
situations and is not restricted to Brownian Dynamics. 

Heterogenous nucleation in a sheared system confined between two
parallel plates was studied experimentally by Palberg and coworkers in
another paper of this special issue~\cite{Stipp:2004JPCM}. A
detailed comparison between experiment and theory is in principle
possible and left for future work. 

\begin{acknowledgments}
We like to thank T. Palberg, A. Van Blaaderen, G. Szamel, and
S. Egelhaaf for helpful discussions. This work has been supported by
the Deutsche Forschungsgemeinschaft (DFG), within subproject D1 of the
SFB-TR6 program. The work of FOM Institute is part of the research
program of the ``Stichting voor Fundamenteel Onderzoek der Materie''
(FOM), which is financially supported by the ``Nederlandse organisatie
voor Wetenschappelijk Onderzoek'' (NWO). 
\end{acknowledgments}


\newpage
\begin{thebibliography}{39}
\expandafter\ifx\csname natexlab\endcsname\relax\def\natexlab#1{#1}\fi
\expandafter\ifx\csname bibnamefont\endcsname\relax
  \def\bibnamefont#1{#1}\fi
\expandafter\ifx\csname bibfnamefont\endcsname\relax
  \def\bibfnamefont#1{#1}\fi
\expandafter\ifx\csname citenamefont\endcsname\relax
  \def\citenamefont#1{#1}\fi
\expandafter\ifx\csname url\endcsname\relax
  \def\url#1{\texttt{#1}}\fi
\expandafter\ifx\csname urlprefix\endcsname\relax\def\urlprefix{URL }\fi
\providecommand{\bibinfo}[2]{#2}
\providecommand{\eprint}[2][]{\url{#2}}

\bibitem[{\citenamefont{Volmer and Weber}(1926)}]{Volmer:1926ZPC}
\bibinfo{author}{\bibfnamefont{M.}~\bibnamefont{Volmer}} \bibnamefont{and}
  \bibinfo{author}{\bibfnamefont{A.}~\bibnamefont{Weber}}, \bibinfo{journal}{Z.
  Phys. Chem.} \textbf{\bibinfo{volume}{119}}, \bibinfo{pages}{277}
  (\bibinfo{year}{1926}).

\bibitem[{\citenamefont{Becker and D\"oring}(1935)}]{Becker:1935AnP}
\bibinfo{author}{\bibfnamefont{R.}~\bibnamefont{Becker}} \bibnamefont{and}
  \bibinfo{author}{\bibfnamefont{W.}~\bibnamefont{D\"oring}},
  \bibinfo{journal}{Ann. Physik} \textbf{\bibinfo{volume}{24}},
  \bibinfo{pages}{719} (\bibinfo{year}{1935}).

\bibitem[{\citenamefont{Kelton}(1991)}]{Kelton:1991}
\bibinfo{author}{\bibfnamefont{K.~F.} \bibnamefont{Kelton}}, in
  \emph{\bibinfo{booktitle}{Solid State Physics}}, edited by
  \bibinfo{editor}{\bibfnamefont{H.}~\bibnamefont{Ehrenreich}}
  \bibnamefont{and} \bibinfo{editor}{\bibfnamefont{D.}~\bibnamefont{Turnbull}}
  (\bibinfo{publisher}{Academic Press}, \bibinfo{address}{New York},
  \bibinfo{year}{1991}), vol.~\bibinfo{volume}{45}, pp.
  \bibinfo{pages}{75--178}.

\bibitem[{\citenamefont{Galkin and Vekilov}(2000)}]{Galkin:2000PNAS}
\bibinfo{author}{\bibfnamefont{O.}~\bibnamefont{Galkin}} \bibnamefont{and}
  \bibinfo{author}{\bibfnamefont{P.~G.} \bibnamefont{Vekilov}},
  \bibinfo{journal}{Proc. Natl. Acad. Sci. USA} \textbf{\bibinfo{volume}{97}},
  \bibinfo{pages}{6277} (\bibinfo{year}{2000}).

\bibitem[{\citenamefont{Shi et~al.}(1995)\citenamefont{Shi, Tong, and
  Ayers}}]{Shi:1995APL}
\bibinfo{author}{\bibfnamefont{F.~G.} \bibnamefont{Shi}},
  \bibinfo{author}{\bibfnamefont{H.~Y.} \bibnamefont{Tong}}, \bibnamefont{and}
  \bibinfo{author}{\bibfnamefont{J.~D.} \bibnamefont{Ayers}},
  \bibinfo{journal}{Appl. Phys. Lett.} \textbf{\bibinfo{volume}{67}},
  \bibinfo{pages}{350} (\bibinfo{year}{1995}).

\bibitem[{\citenamefont{Sch\"atzel and Ackerson}(1993)}]{Schatzel:1993PRE}
\bibinfo{author}{\bibfnamefont{K.}~\bibnamefont{Sch\"atzel}} \bibnamefont{and}
  \bibinfo{author}{\bibfnamefont{B.~J.} \bibnamefont{Ackerson}},
  \bibinfo{journal}{Phys. Rev. E} \textbf{\bibinfo{volume}{48}},
  \bibinfo{pages}{3766} (\bibinfo{year}{1993}).

\bibitem[{\citenamefont{Harland and {van Megen}}(1997)}]{Harland:1997PRE}
\bibinfo{author}{\bibfnamefont{J.~L.} \bibnamefont{Harland}} \bibnamefont{and}
  \bibinfo{author}{\bibfnamefont{W.}~\bibnamefont{{van Megen}}},
  \bibinfo{journal}{Phys. Rev. E} \textbf{\bibinfo{volume}{55}},
  \bibinfo{pages}{3054} (\bibinfo{year}{1997}).

\bibitem[{\citenamefont{Sinn et~al.}(2001)\citenamefont{Sinn, Heymann, Stipp,
  and Palberg}}]{Sinn:2001PCPS}
\bibinfo{author}{\bibfnamefont{C.}~\bibnamefont{Sinn}},
  \bibinfo{author}{\bibfnamefont{A.}~\bibnamefont{Heymann}},
  \bibinfo{author}{\bibfnamefont{A.}~\bibnamefont{Stipp}}, \bibnamefont{and}
  \bibinfo{author}{\bibfnamefont{T.}~\bibnamefont{Palberg}},
  \bibinfo{journal}{Prog. Colloid Polymer Sci.} \textbf{\bibinfo{volume}{118}},
  \bibinfo{pages}{266} (\bibinfo{year}{2001}).

\bibitem[{\citenamefont{Pusey}(1991)}]{Pusey:1991}
\bibinfo{author}{\bibfnamefont{P.}~\bibnamefont{Pusey}}, in
  \emph{\bibinfo{booktitle}{Liquids, Freezing and the Glass Transition}},
  edited by \bibinfo{editor}{\bibfnamefont{J.~P.} \bibnamefont{Hansen}},
  \bibinfo{editor}{\bibfnamefont{D.}~\bibnamefont{Levesque}}, \bibnamefont{and}
  \bibinfo{editor}{\bibfnamefont{J.}~\bibnamefont{Zinn-Justin}}
  (\bibinfo{publisher}{North-Holland}, \bibinfo{address}{Amsterdam},
  \bibinfo{year}{1991}), pp. \bibinfo{pages}{763--942}.

\bibitem[{\citenamefont{Gasser et~al.}(2001)\citenamefont{Gasser, Weeks,
  Schofield, Pusey, and Weitz}}]{Gasser:2001SC}
\bibinfo{author}{\bibfnamefont{U.}~\bibnamefont{Gasser}},
  \bibinfo{author}{\bibfnamefont{E.~R.} \bibnamefont{Weeks}},
  \bibinfo{author}{\bibfnamefont{A.}~\bibnamefont{Schofield}},
  \bibinfo{author}{\bibfnamefont{P.~N.} \bibnamefont{Pusey}}, \bibnamefont{and}
  \bibinfo{author}{\bibfnamefont{D.~A.} \bibnamefont{Weitz}},
  \bibinfo{journal}{Science} \textbf{\bibinfo{volume}{292}},
  \bibinfo{pages}{258} (\bibinfo{year}{2001}).

\bibitem[{\citenamefont{Auer and Frenkel}(2001)}]{Auer:2001NAT1}
\bibinfo{author}{\bibfnamefont{S.}~\bibnamefont{Auer}} \bibnamefont{and}
  \bibinfo{author}{\bibfnamefont{D.}~\bibnamefont{Frenkel}},
  \bibinfo{journal}{Nature} \textbf{\bibinfo{volume}{409}},
  \bibinfo{pages}{1020} (\bibinfo{year}{2001}).

\bibitem[{\citenamefont{Auer and Frenkel}(2002)}]{Auer:2002JPCM}
\bibinfo{author}{\bibfnamefont{S.}~\bibnamefont{Auer}} \bibnamefont{and}
  \bibinfo{author}{\bibfnamefont{D.}~\bibnamefont{Frenkel}},
  \bibinfo{journal}{J. Phys.: Condens. Matter} \textbf{\bibinfo{volume}{14}},
  \bibinfo{pages}{7667} (\bibinfo{year}{2002}).

\bibitem[{\citenamefont{Ackerson and Pusey}(1988)}]{Ackerson:1988PRL}
\bibinfo{author}{\bibfnamefont{B.~J.} \bibnamefont{Ackerson}} \bibnamefont{and}
  \bibinfo{author}{\bibfnamefont{P.~N.} \bibnamefont{Pusey}},
  \bibinfo{journal}{Phys. Rev. Lett.} \textbf{\bibinfo{volume}{61}},
  \bibinfo{pages}{1033} (\bibinfo{year}{1988}).

\bibitem[{\citenamefont{Yan et~al.}(1994)\citenamefont{Yan, Dhont, Smits, and
  Lekkerkerker}}]{Yan:1994PA}
\bibinfo{author}{\bibfnamefont{Y.~D.} \bibnamefont{Yan}},
  \bibinfo{author}{\bibfnamefont{J.~K.~G.} \bibnamefont{Dhont}},
  \bibinfo{author}{\bibfnamefont{C.}~\bibnamefont{Smits}}, \bibnamefont{and}
  \bibinfo{author}{\bibfnamefont{H.~N.~W.} \bibnamefont{Lekkerkerker}},
  \bibinfo{journal}{Physica A} \textbf{\bibinfo{volume}{202}},
  \bibinfo{pages}{68} (\bibinfo{year}{1994}).

\bibitem[{\citenamefont{Haw et~al.}(1998)\citenamefont{Haw, Poon, and
  Pusey}}]{Haw:1998PRE}
\bibinfo{author}{\bibfnamefont{M.~D.} \bibnamefont{Haw}},
  \bibinfo{author}{\bibfnamefont{W.~C.~K.} \bibnamefont{Poon}},
  \bibnamefont{and} \bibinfo{author}{\bibfnamefont{P.~N.} \bibnamefont{Pusey}},
  \bibinfo{journal}{Phys. Rev. E} \textbf{\bibinfo{volume}{57}},
  \bibinfo{pages}{6859} (\bibinfo{year}{1998}).

\bibitem[{\citenamefont{Amos et~al.}(2000)\citenamefont{Amos, Rarity, Tapster,
  Shepherd, and Kitson}}]{Amos:2000PRE}
\bibinfo{author}{\bibfnamefont{R.~M.} \bibnamefont{Amos}},
  \bibinfo{author}{\bibfnamefont{J.~G.} \bibnamefont{Rarity}},
  \bibinfo{author}{\bibfnamefont{P.~R.} \bibnamefont{Tapster}},
  \bibinfo{author}{\bibfnamefont{T.~J.} \bibnamefont{Shepherd}},
  \bibnamefont{and} \bibinfo{author}{\bibfnamefont{S.~C.}
  \bibnamefont{Kitson}}, \bibinfo{journal}{Phys. Rev. E}
  \textbf{\bibinfo{volume}{61}}, \bibinfo{pages}{2929} (\bibinfo{year}{2000}).

\bibitem[{\citenamefont{Palberg et~al.}(1995)\citenamefont{Palberg, M\"onch,
  Schwarz, and Leiderer}}]{Palberg:1995JCP}
\bibinfo{author}{\bibfnamefont{T.}~\bibnamefont{Palberg}},
  \bibinfo{author}{\bibfnamefont{W.}~\bibnamefont{M\"onch}},
  \bibinfo{author}{\bibfnamefont{J.}~\bibnamefont{Schwarz}}, \bibnamefont{and}
  \bibinfo{author}{\bibfnamefont{P.}~\bibnamefont{Leiderer}},
  \bibinfo{journal}{J. Chem. Phys.} \textbf{\bibinfo{volume}{102}},
  \bibinfo{pages}{5082} (\bibinfo{year}{1995}).

\bibitem[{\citenamefont{Okubo and Ishiki}(1999)}]{Okubo:1999JCIS}
\bibinfo{author}{\bibfnamefont{T.}~\bibnamefont{Okubo}} \bibnamefont{and}
  \bibinfo{author}{\bibfnamefont{H.}~\bibnamefont{Ishiki}},
  \bibinfo{journal}{J. Colloid and Interface Science}
  \textbf{\bibinfo{volume}{211}}, \bibinfo{pages}{151} (\bibinfo{year}{1999}).

\bibitem[{\citenamefont{Stipp et~al.}(2004)\citenamefont{Stipp, Biehl, Preis,
  Liu, {Barreira Fontecha}, Sch\"ope, and Palberg}}]{Stipp:2004JPCM}
\bibinfo{author}{\bibfnamefont{A.}~\bibnamefont{Stipp}},
  \bibinfo{author}{\bibfnamefont{R.}~\bibnamefont{Biehl}},
  \bibinfo{author}{\bibfnamefont{T.}~\bibnamefont{Preis}},
  \bibinfo{author}{\bibfnamefont{J.}~\bibnamefont{Liu}},
  \bibinfo{author}{\bibfnamefont{A.}~\bibnamefont{{Barreira Fontecha}}},
  \bibinfo{author}{\bibfnamefont{H.~J.} \bibnamefont{Sch\"ope}},
  \bibnamefont{and} \bibinfo{author}{\bibfnamefont{T.}~\bibnamefont{Palberg}},
  \bibinfo{journal}{J. Phys.: Condens. Matter} \textbf{\bibinfo{volume}{this
  spececial isue}} (\bibinfo{year}{2004}).

\bibitem[{\citenamefont{Butler and Harrowell}(1995)}]{Butler:1995PRE}
\bibinfo{author}{\bibfnamefont{S.}~\bibnamefont{Butler}} \bibnamefont{and}
  \bibinfo{author}{\bibfnamefont{P.}~\bibnamefont{Harrowell}},
  \bibinfo{journal}{Phys. Rev. E} \textbf{\bibinfo{volume}{52}},
  \bibinfo{pages}{6424} (\bibinfo{year}{1995}).

\bibitem[{\citenamefont{Blaak et~al.}(to be published 2004)\citenamefont{Blaak,
  Auer, Frenkel, and L\"owen}}]{Blaak:2004PRL}
\bibinfo{author}{\bibfnamefont{R.}~\bibnamefont{Blaak}},
  \bibinfo{author}{\bibfnamefont{S.}~\bibnamefont{Auer}},
  \bibinfo{author}{\bibfnamefont{D.}~\bibnamefont{Frenkel}}, \bibnamefont{and}
  \bibinfo{author}{\bibfnamefont{H.}~\bibnamefont{L\"owen}},
  \bibinfo{journal}{to appear in Phys. Rev. Lett.}  (cond-mat/0406237).

\bibitem[{\citenamefont{Allen and Tildesley}(1987)}]{Book:Allen-Tildesley}
\bibinfo{author}{\bibfnamefont{M.~P.} \bibnamefont{Allen}} \bibnamefont{and}
  \bibinfo{author}{\bibfnamefont{D.~J.} \bibnamefont{Tildesley}},
  \emph{\bibinfo{title}{Computer simulations of liquids}}
  (\bibinfo{publisher}{Oxford University Press}, \bibinfo{address}{Oxford},
  \bibinfo{year}{1987}).

\bibitem[{\citenamefont{Chakrabarti et~al.}(1994)\citenamefont{Chakrabarti,
  Sood, and Krishnamurthy}}]{Chakrabarti:1994PRE}
\bibinfo{author}{\bibfnamefont{J.}~\bibnamefont{Chakrabarti}},
  \bibinfo{author}{\bibfnamefont{A.~K.} \bibnamefont{Sood}}, \bibnamefont{and}
  \bibinfo{author}{\bibfnamefont{H.~R.} \bibnamefont{Krishnamurthy}},
  \bibinfo{journal}{Phys. Rev. E} \textbf{\bibinfo{volume}{50}},
  \bibinfo{pages}{R3326} (\bibinfo{year}{1994}).

\bibitem[{\citenamefont{Lees and Edwards}(1972)}]{Lees:1972JPCS}
\bibinfo{author}{\bibfnamefont{A.~W.} \bibnamefont{Lees}} \bibnamefont{and}
  \bibinfo{author}{\bibfnamefont{S.~F.} \bibnamefont{Edwards}},
  \bibinfo{journal}{J. Phys. C} \textbf{\bibinfo{volume}{5}},
  \bibinfo{pages}{1921} (\bibinfo{year}{1972}).

\bibitem[{\citenamefont{Steinhardt et~al.}(1983)\citenamefont{Steinhardt,
  Nelson, and Ronchetti}}]{Steinhardt:1983PRB}
\bibinfo{author}{\bibfnamefont{P.~J.} \bibnamefont{Steinhardt}},
  \bibinfo{author}{\bibfnamefont{D.~R.} \bibnamefont{Nelson}},
  \bibnamefont{and}
  \bibinfo{author}{\bibfnamefont{M.}~\bibnamefont{Ronchetti}},
  \bibinfo{journal}{Phys. Rev. B} \textbf{\bibinfo{volume}{28}},
  \bibinfo{pages}{784} (\bibinfo{year}{1983}).

\bibitem[{\citenamefont{{van Duijneveldt} and
  Frenkel}(1992)}]{Duijneveldt:1992JCP}
\bibinfo{author}{\bibfnamefont{J.~S.} \bibnamefont{{van Duijneveldt}}}
  \bibnamefont{and} \bibinfo{author}{\bibfnamefont{D.}~\bibnamefont{Frenkel}},
  \bibinfo{journal}{J. Chem. Phys.} \textbf{\bibinfo{volume}{96}},
  \bibinfo{pages}{4655} (\bibinfo{year}{1992}).

\bibitem[{\citenamefont{{ten Wolde} et~al.}(1995)\citenamefont{{ten Wolde},
  Ruiz-Montero, and Frenkel}}]{Wolde:1995PRL}
\bibinfo{author}{\bibfnamefont{P.~R.} \bibnamefont{{ten Wolde}}},
  \bibinfo{author}{\bibfnamefont{M.~J.} \bibnamefont{Ruiz-Montero}},
  \bibnamefont{and} \bibinfo{author}{\bibfnamefont{D.}~\bibnamefont{Frenkel}},
  \bibinfo{journal}{Phys. Rev. Lett.} \textbf{\bibinfo{volume}{75}},
  \bibinfo{pages}{2714} (\bibinfo{year}{1995}).

\bibitem[{\citenamefont{Auer and Frenkel}(2004)}]{Auer:2004JCP}
\bibinfo{author}{\bibfnamefont{S.}~\bibnamefont{Auer}} \bibnamefont{and}
  \bibinfo{author}{\bibfnamefont{D.}~\bibnamefont{Frenkel}},
  \bibinfo{journal}{J. Chem. Phys.} \textbf{\bibinfo{volume}{120}},
  \bibinfo{pages}{3015} (\bibinfo{year}{2004}).

\bibitem[{\citenamefont{O'Malley and Snook}(2003)}]{OMalley:2003PRL}
\bibinfo{author}{\bibfnamefont{B.}~\bibnamefont{O'Malley}} \bibnamefont{and}
  \bibinfo{author}{\bibfnamefont{I.}~\bibnamefont{Snook}},
  \bibinfo{journal}{Phys. Rev. Lett.} \textbf{\bibinfo{volume}{90}},
  \bibinfo{pages}{085702} (\bibinfo{year}{2003}).

\bibitem[{\citenamefont{Torrie and Valleau}(1974)}]{Torrie:1974CPL}
\bibinfo{author}{\bibfnamefont{G.~M.} \bibnamefont{Torrie}} \bibnamefont{and}
  \bibinfo{author}{\bibfnamefont{J.~P.} \bibnamefont{Valleau}},
  \bibinfo{journal}{Chem. Phys. Lett.} \textbf{\bibinfo{volume}{28}},
  \bibinfo{pages}{578} (\bibinfo{year}{1974}).

\bibitem[{\citenamefont{Hamaguchi et~al.}(1997)\citenamefont{Hamaguchi,
  Farouki, and Dubin}}]{Hamaguchi:1997PRE}
\bibinfo{author}{\bibfnamefont{S.}~\bibnamefont{Hamaguchi}},
  \bibinfo{author}{\bibfnamefont{R.~T.} \bibnamefont{Farouki}},
  \bibnamefont{and} \bibinfo{author}{\bibfnamefont{D.~H.~E.}
  \bibnamefont{Dubin}}, \bibinfo{journal}{Phys. Rev. E}
  \textbf{\bibinfo{volume}{56}}, \bibinfo{pages}{4671} (\bibinfo{year}{1997}).

\bibitem[{\citenamefont{Daivis and Matin}(2003)}]{Daivis:2003JCP}
\bibinfo{author}{\bibfnamefont{P.~J.} \bibnamefont{Daivis}} \bibnamefont{and}
  \bibinfo{author}{\bibfnamefont{M.~L.} \bibnamefont{Matin}},
  \bibinfo{journal}{J. Chem. Phys.} \textbf{\bibinfo{volume}{118}},
  \bibinfo{pages}{11111} (\bibinfo{year}{2003}).

\bibitem[{\citenamefont{Stevens and Robbins}(1993)}]{Stevens:1993PRE}
\bibinfo{author}{\bibfnamefont{M.~J.} \bibnamefont{Stevens}} \bibnamefont{and}
  \bibinfo{author}{\bibfnamefont{M.~O.} \bibnamefont{Robbins}},
  \bibinfo{journal}{Phys. Rev. E} \textbf{\bibinfo{volume}{48}},
  \bibinfo{pages}{3778} (\bibinfo{year}{1993}).

\bibitem[{\citenamefont{Butler and Harrowell}(2003)}]{Butler:2003JCP}
\bibinfo{author}{\bibfnamefont{S.}~\bibnamefont{Butler}} \bibnamefont{and}
  \bibinfo{author}{\bibfnamefont{P.}~\bibnamefont{Harrowell}},
  \bibinfo{journal}{J. Chem. Phys.} \textbf{\bibinfo{volume}{118}},
  \bibinfo{pages}{4115} (\bibinfo{year}{2003}).

\bibitem[{\citenamefont{Abkarian et~al.}(2002)\citenamefont{Abkarian, Lartigue,
  and Viallat}}]{Abkarian:2002PRL}
\bibinfo{author}{\bibfnamefont{M.}~\bibnamefont{Abkarian}},
  \bibinfo{author}{\bibfnamefont{C.}~\bibnamefont{Lartigue}}, \bibnamefont{and}
  \bibinfo{author}{\bibfnamefont{A.}~\bibnamefont{Viallat}},
  \bibinfo{journal}{Phys. Rev. Lett.} \textbf{\bibinfo{volume}{88}},
  \bibinfo{pages}{068103} (\bibinfo{year}{2002}).

\bibitem[{\citenamefont{Palberg and W\"urth}(1996)}]{Palberg:1996JP1F}
\bibinfo{author}{\bibfnamefont{T.}~\bibnamefont{Palberg}} \bibnamefont{and}
  \bibinfo{author}{\bibfnamefont{M.}~\bibnamefont{W\"urth}},
  \bibinfo{journal}{J. Phys. I France} \textbf{\bibinfo{volume}{6}},
  \bibinfo{pages}{237} (\bibinfo{year}{1996}).

\bibitem[{\citenamefont{Reguera and Rubi}(2003)}]{Reguera:2003JCP2}
\bibinfo{author}{\bibfnamefont{D.}~\bibnamefont{Reguera}} \bibnamefont{and}
  \bibinfo{author}{\bibfnamefont{J.~M.} \bibnamefont{Rubi}},
  \bibinfo{journal}{J. Chem. Phys.} \textbf{\bibinfo{volume}{119}},
  \bibinfo{pages}{9888} (\bibinfo{year}{2003}).

\bibitem[{\citenamefont{Xue and Grest}(1989)}]{Xue:1989PRA}
\bibinfo{author}{\bibfnamefont{W.}~\bibnamefont{Xue}} \bibnamefont{and}
  \bibinfo{author}{\bibfnamefont{G.~S.} \bibnamefont{Grest}},
  \bibinfo{journal}{Phys. Rev. A} \textbf{\bibinfo{volume}{40}},
  \bibinfo{pages}{R1709} (\bibinfo{year}{1989}).

\bibitem[{\citenamefont{Auer and Frenkel}(2003)}]{Auer:2003PRL}
\bibinfo{author}{\bibfnamefont{S.}~\bibnamefont{Auer}} \bibnamefont{and}
  \bibinfo{author}{\bibfnamefont{D.}~\bibnamefont{Frenkel}},
  \bibinfo{journal}{Phys. Rev. Lett.} \textbf{\bibinfo{volume}{91}},
  \bibinfo{pages}{015703} (\bibinfo{year}{2003}).

\end{thebibliography}
\end{document}